\documentstyle[12pt]{article}

\def\+{{+\!\!\!+}} 
\def\pp{\mbox{\tiny${}_{\stackrel\+ =}$}}

\def\d{\partial}

\def\th{\theta}

\def\P{\Phi}

\def\p{\psi} 
 
\def\e{\varepsilon}

\def\pmb#1{\setbox0=\hbox{#1}%
\kern.0em\copy0\kern-\wd0 
\kern-.04em\copy0\kern-\wd0 
\kern.08em\copy0\kern-\wd0 
\kern-.04em\raise.0433em\box0 }         
 
\newcommand{\nc}{\newcommand} 
\nc{\beq}{\begin{equation}} 
\nc{\eeq}[1]{\label{#1}\end{equation}} 
\nc{\ber}{\begin{eqnarray}} 
\nc{\eer}[1]{\label{#1}\end{eqnarray}} 
\nc{\pek}[1]{\cite{#1}} 
\nc{\enr}[1]{(\ref{#1})} 
\nc{\kal}[1]{{\cal{#1}}} 
\nc{\dott}{\;\cdot\;} 
\newcommand{\Section}[1]{\section{#1} \setcounter{equation}{0}} 
 
\begin{document} 
\newcommand{\inv}[1]{{#1}^{-1}} 
\renewcommand{\theequation}{\thesection.\arabic{equation}} 
\newcommand{\be}{\begin{equation}} 
\newcommand{\ee}{\end{equation}} 
\newcommand{\bea}{\begin{eqnarray}} 
\newcommand{\eea}{\end{eqnarray}} 
\newcommand{\re}[1]{(\ref{#1})} 
\newcommand{\qv}{\quad ,} 
\newcommand{\qp}{\quad .} 
\begin{center} 
                                \hfill UUITP-18/02\\
                                \hfill   hep-th/0212042\\ 
\vskip .3in \noindent 
 
\vskip .1in 
 
{\large \bf{D-branes in N=2 WZW models}} 
\vskip .2in 
 
{\bf Ulf Lindstr\"om}$^a$\footnote{e-mail address: ulf.lindstrom@teorfys.uu.se} 
 and  {\bf Maxim Zabzine}$^{b}$\footnote{e-mail address: zabzine@fi.infn.it} \\ 
 
 
\vskip .15in 
 
\vskip .15in 
$^a${\em Department of Theoretical Physics, \\
Uppsala University,
Box 803, SE-751 08 Uppsala, Sweden }\\ 
\vskip .15in 
$^b${\em INFN Sezione di Firenze, Dipartimento di Fisica\\ 
 Via Sansone 1, I-50019 Sesto F.no (FI), Italy} 
 
\bigskip 
 
 
 \vskip .1in 
\end{center} 
\vskip .4in 
 
\begin{center} {\bf ABSTRACT }  
\end{center} 
\begin{quotation}\noindent  
 We briefly review the construction of N=2 WZW models in terms of Manin
 triples. We analyse the restrictions which should be imposed on 
 the gluing conditions of the affine currents in order to preserve 
  half of the bulk supersymmetry. 
 In analogy with the K\"ahler case there are two types of D-branes, A- and B-types
 which have a nice algebraic 
 interpretation in terms of the Manin triple.
\end{quotation} 
\vfill 
\eject 
  
 
\section{Introduction}

 During the last few years D-branes on group manifolds have received a great deal of attention
 (see, e.g., \cite{Schomerus:2002dc} and the references therein).
  Using methods from CFT one obtains a microscopic decription of various D-branes on group
  manifolds in terms of conformally invariant boundary states.
 Therefore  the group manifolds provide an ideal laboratory for the study of 
 quantum D-branes on general curved backgrounds. 
 
Recently, the study of boundary conditions that preserve various symmetries
 on the boundary of sigma models representing open strings
\cite{Haggi-Mani:2000uc}-\cite{Lindstrom:2002mc} 
has revealed a close connection to the geometry of the targetspace.
 The present letter may be viewed as an application of results from those studies.

 Here we study D-branes on group manifolds which 
 preserve a certain amount of world-sheet supersymmetry, in particular we focus on  the N=2 case.
   It turns out that for the WZW model N=2 supersymmetry alone imposes strong restrictions
  on the possible  gluing conditions of the affine currents and that these restrictions have a nice
 algebraic description.  The study in this letter is confined to the case
 where the gluing conditions imposed on the affine currents are given in terms
 of a constant map $R^A_{\,\,B}$. Although this restriction is not necessary, the case of
 constant $R^A_{\,\,B}$ is the most interesting from the CFT point of view.
 In  this study we apply our previous results \cite{Lindstrom:2002jb} for D-branes
 of general N=(2,2) supersymmetric sigma model to the WZW model.

The letter is organised as follows. In Section~\ref{s:review} we review the N=(2,2) supersymmetry 
 for the WZW models and briefly explain how N=2 supersymmetry is related to the Manin triple $({\bf g},
 {\bf g_-}, {\bf g_+})$. In Section~\ref{s:susy} using the N=2 on-shell transformations 
  we study the requirements which need to be imposed on the gluing conditions of the affine currents
  to preserve half of the world-sheet supersymmetry. In analogy with the K\"ahler case
 we introduce two type of branes: A-type and B-type. Both types of branes
 have an algebraic description in terms of the Manin triple. In Section~\ref{s:conformal}
 we analyse the N=2 superconformal boundary conditions by imposing
 boundary conditions on the N=2 currents. Finally, in Section~\ref{s:end}, we summarize the 
 results and give some examples.   

\section{N=2 WZW models} 
\label{s:review}

 First, let us recall the description of N=2 supersymmetry for non-linear 
 sigma models with torsion \cite{Gates:nk} (for a recent discussion
  see \cite{Lyakhovich:2002kc}). 
 The N=1 superfield bulk action for the real scalar superfields
 $\Phi^\mu$ is
\beq
 S= \int d^2\sigma\,d^2\theta\,\,D_+\Phi^\mu D_- \Phi^\nu (g_{\mu\nu}(\Phi) 
 + B_{\mu\nu}(\Phi)) ,
\eeq{actionB}
 where we assume that $H\equiv dB\neq 0$. This action is manifestly supersymmetric
 under one supersymmetry because of its N=1 superfield form. Further (\ref{actionB})
 admits an additional nonmanifest supersymmetry
 of the form
\beq
 \delta_2 \Phi^\mu =  \epsilon_2^+ D_+ \Phi^\nu J^\mu_{+\nu}(\Phi)
  + \epsilon_2^- D_- \Phi^\nu J^\mu_{-\nu}(\Phi) . 
\eeq{secsupfl}
  Classically this ansatz is unique for dimensional reason. The standard on-shell N=2
 supersymmetry requires
 $J^\mu_{\pm\nu}$ to be complex structures, i.e.
\beq
 J^\mu_{\pm\lambda} J^\lambda_{\pm\nu} = - \delta^\mu_{\,\,\nu}
\eeq{susystract}
 and 
\beq
  {\cal N}^{\rho}_{\,\,\mu\nu} (J_{\pm}) = J^\gamma_{\pm\mu} \d_{[\gamma} J^\rho_{\pm\nu]}
 - J^\gamma_{\pm\nu} \d_{[\gamma} J^\rho_{\pm\mu]}=0 .
\eeq{vanishNT}
 Invariance of the action (\ref{actionB}) under the transformations (\ref{secsupfl}) 
 requires the metric $g_{\mu\nu}$ to be Hermitian
 with respect to both complex structures
\beq
 J^\mu_{\pm\rho} g_{\mu\nu} J^\nu_{\pm\lambda} = g_{\rho\lambda}
\eeq{herm}
 and the complex structures to be covariantly constant with respect 
 to different connections 
\beq
 \nabla^{(\pm)}_\rho J^\mu_{\pm\nu} \equiv J^\mu_{\pm\nu,\rho} +
 \Gamma^{\pm\mu}_{\,\,\rho\sigma} J^\sigma_{\pm\nu} - \Gamma^{\pm\sigma}_{\,\,\rho\nu}
 J^\mu_{\pm\sigma}=0 ,
\eeq{nablJH}  
 where we have defined the two affine connections as 
\beq
 \Gamma^{\pm\mu}_{\,\,\rho\nu} = \Gamma^{\mu}_{\,\,\rho\nu} \pm g^{\mu\sigma} H_{\sigma\rho\nu},
\eeq{defaffcon}
with $\Gamma^{\mu}_{\,\,\rho\nu}$ the Christoffel connection for the metric $g_{\mu\nu}$. Using (\ref{nablJH}) the inegrability condition (\ref{vanishNT}) 
 may be rewritten in alternative form
 \beq
 H_{\delta\nu\lambda} = J^\sigma_{\pm\delta} J^\rho_{\pm\nu} H_{\sigma\rho\lambda} +
 J^\sigma_{\pm\lambda} J^\rho_{\pm\delta} H_{\sigma\rho\nu}+
 J^\sigma_{\pm\nu} J^\rho_{\pm\lambda} H_{\sigma\rho\delta} . 
\eeq{inegrabtroB}
 To summarize, the transformation (\ref{secsupfl}) is a second supersymmetry provided that 
  (\ref{susystract}), (\ref{herm}), (\ref{nablJH}) and (\ref{inegrabtroB})
  are satisfied. 

 We now turn to N=2 supersymmetric WZW models. 
 The WZW models represent a special class of non-linear sigma models defined over a group
 manifold ${\cal M}$ of some Lie group ${\cal G}$. The isometry group, ${\cal G}\times
 {\cal G}$  is generated by the left and right  invariant Killing vectors $l^\mu_A$ and 
  $r^\mu_A$ respectively, where $A=1,2,...,\dim\,{\cal G}$. They satisfy
\beq
 \{ l_A, l_B \} = f_{AB}^{\,\,\,\,\,\,\,\,C} l_C,\,\,\,\,\,\,\,\,\,\,\,\,
 \{ r_A, r_B \} = -f_{AB}^{\,\,\,\,\,\,\,\,C} r_C,\,\,\,\,\,\,\,\,\,\,\,\,
 \{ l_A, r_B\}=0 ,
\eeq{liebrakalg}
 where $\{,\}$ is the Lie bracket for the vector fields. 
 We restrict ourselves to semi-simple
 Lie groups, so that the Cartan-Killing metric $\eta_{AB}$ has an inverse $\eta^{AB}$ and can 
 be used to raise and lower Lie algebra indices. Both $l^\mu_A$ and $r^\mu_A$ can be regarded
 as veilbeins, with inverses $l^A_\mu$, $r^A_\mu$ respectively. To define the sigma 
 model, we choose the invariant metric
\beq
 g_{\mu\nu} = \frac{1}{\rho^2}\, l^A_\mu l^B_\nu \eta_{AB} =
 \frac{1}{\rho^2}\, r^A_\mu r^B_\nu \eta_{AB} 
\eeq{metrinv}
 while $H_{\mu\nu\rho}$ is proportional to the structure constants of the corresponding 
 Lie algebra ${\bf g}$
\beq
 H_{\mu\nu\rho} = \frac{1}{2} k\, l^A_\mu l^B_\nu l^C_\rho f_{ABC} =
 \frac{1}{2} k\, r^A_\mu r^B_\nu r^C_\rho f_{ABC} 
\eeq{deftors} 
 and  where $\rho$ and $k$ are constants and $k$ must satisfy a quantization condition. 
  If $\rho^2 = \pm 1/k$ then $H_{\mu\nu\rho}$ is the parallelizing torsion on the group 
 manifold and this is also precisely the relation bewteen the coupling constants that holds
  at the conformally invariant fixed-point of the beta-functions. Since we are interested in 
 the conformal model, in the following discussion
 we set $\rho^2=1/k$ and $k=1$ because in our calculations $k$ appears only as overall factor. 
 We are thus interested in the sigma model (\ref{actionB}) with $g_{\mu\nu}$ and $H_{\mu\nu\rho}$
 given by (\ref{metrinv}) and (\ref{deftors}). Using the above properties we see that the 
 left and right invariant Killing vectors satisfy the following equations
\beq
\nabla^{(-)}_\rho l^\mu_A =0,\,\,\,\,\,\,\,\,\,\,\,\,\,\,\,\,\,\,\nabla^{(+)}_\rho r^\mu_A=0,
\eeq{covconKV}
 where $\nabla_\rho^{(\pm)}$ are the affine connections defined in (\ref{defaffcon}). 
 Due to (\ref{covconKV}) there are chiral (antichiral) Lie algebra valued currents
\beq
 {\cal J}^A_- = l^A_\mu D_- \Phi^\mu,\,\,\,\,\,\,\,\,\,\,\,\,\,\,
 {\cal J}^A_+ = r^A_\mu D_+ \Phi^\mu,
\eeq{fefgroupcur}
 such that $D_{\mp} {\cal J}^A_{\pm} =0$. The components of these currents are defined 
 as follows
\beq
 j_{\pm}^A = {\cal J}^A_{\pm}|,\,\,\,\,\,\,\,\,\,\,\,\,\,\,\,\,
 k^A_{\pp} = -iD_{\pm} {\cal J}^A_{\pm}| .
\eeq{compgrcur}
 Instead of using coordinates $\Phi^\mu$, the group manifold can be parametrized by group
 elements in some representation with the generators $T_A$ satisfying
\beq
 [T_A, T_B] =  f_{AB}^{\,\,\,\,\,\,\,\,C} T_C,\,\,\,\,\,\,\,\,\,\,\,\,\,\,
 tr(T_A T_B) = - \kappa \eta_{AB} .
\eeq{groprepr}
 However in this letter we will not use the parametrization in terms
  of group elements. 

 The problem of N=2 supersymmetry for the WZW models was first addressed in
  \cite{Spindel:1988nh}-\cite{Sevrin:1988ps}.
 However, we will not follow the original presentation. 

 Let us assume that the complex structures in (\ref{secsupfl}) have the
 form
\beq
 J^\mu_{-\nu} = l^\mu_A J^A_{\,\,B} l^B_\nu,\,\,\,\,\,\,\,\,\,\,\,\,\,\,\,
 J^\mu_{+\nu} = r^\mu_A J^A_{\,\,B} r^B_\nu
\eeq{supersymemN2c}
 where $J^A_{\,\,B}$ is a constant matrix acting on the Lie algebra. The relations 
 (\ref{nablJH}) are then  automatically satisfied. The remaining properties
(\ref{susystract}), (\ref{herm}) and (\ref{inegrabtroB}) may be rewritten in terms of 
 $J^A_{\,\,B}$ as follows 
\beq
 J^A_{\,\,C} J^C_{\,\,B}= -\delta^A_{\,\,B}
\eeq{Jgrppro}
\beq
 J^C_{\,\,A} \eta_{CD} J^D_{\,\,B} = \eta_{AB}
\eeq{hemrJAB}
\beq
 f_{ABC} = J^D_{\,\,A} J^L_{\,\,B} f_{DLC} + J^D_{\,\,B} J^L_{\,\,C} f_{DLA}
 + J^D_{\,\,C} J^L_{\,\,A} f_{DLB}
\eeq{HJHHAB}
 Thus we have to construct a $J^A_{\,\,B}$ on the Lie algebra ${\bf g}$ with 
 properties (\ref{Jgrppro})-(\ref{HJHHAB}). That is possible only for even dimensional
 Lie algebras. $J^A_{\,\,B}$ has  as eigenvalues $\pm i$ and we choose a basis 
 $T_A=(T_a,T_{\bar{a}})$
 on the Lie algebra ${\bf g}$ such that $J^A_{\,\,B}$ is diagonal: $J^a_{\,\,b}= i \delta^a_{\,\,b}$, 
  $J^{\bar{a}}_{\,\,\bar{b}}= - i \delta^{\bar{a}}_{\,\,\bar{b}}$. 
 In this basis  eq. (\ref{hemrJAB}) leads to $\eta_{ab}=\eta_{\bar{a}\bar{b}}=0$ 
 and the (\ref{HJHHAB}) gives $f_{abc}=f_{\bar{a}\bar{b}\bar{c}}=0$. 
  Taken together, this implies that $f_{ab}^{\,\,\,\,\,\,{\bar{c}}}=0$
 and $f_{\bar{a}\bar{b}}^{\,\,\,\,\,\,c}=0$. Therefore $\{T_a\}$ and $\{ T_{\bar{a}}\}$
 form Lie subalgebras ${\bf g_+}$ and ${\bf g_-}$ correspondingly. These subalgebras
 are also maximally isotropic subspaces with the respect to $\eta$. Thus the complex
 structures on the even dimensional group is related to a decomposition
 of Lie algebra ${\bf g}$ into two maximally isotropic sublagebras with respect 
 to $\eta$ such that ${\bf g} = {\bf g_-} \oplus {\bf g_+}$ as a vector space.
  Such a structure is called a Manin triple $({\bf g}, {\bf g_-}, {\bf g_+})$
  and  was initially introduced by Drinfeld in the context of completly
 integrable systems and quantum groups \cite{Drinfeld:in}.  
 The relevance of the Manin triples to the N=2 supersymmetry on the group manifolds
 was pointed out  in \cite{Parkhomenko:dq}. 

In the general situation the relation (\ref{supersymemN2c}) should be replaced by the following
\beq
 J^\mu_{-\nu} = l^\mu_A J^A_{\,\,B} l^B_\nu,\,\,\,\,\,\,\,\,\,\,\,\,\,\,\,
 J^\mu_{+\nu} = r^\mu_A \tilde{J}^A_{\,\,B} r^B_\nu
\eeq{supersymemN2cgen}
 where $J^A_{\,\,B}$ and $\tilde{J}^A_{\,\,B}$ should each satisfy the relations 
 (\ref{Jgrppro})-(\ref{HJHHAB}) and thus they should be identified with different
 Manin triples. Therefore, in the general case, the left and right supersymmetries may
 correspond to different Manin triples (however with respect to the same ad-invariant
 bilinear  nondegenerate form $\eta$). Here we only consider the situation 
 where left and right supersymmetries correspond to the same Manin triple.

\Section{N=2 boundary conditions}
\label{s:susy}

In this section we discuss the boundary conditions of N=2 WZW models
  from the point of view of N=2 supersymmetry. In the next section we address
 the N=2 superconformal boundary conditions. 

 Since the classical model does not have a dimensionful parameter
 the most general local classical boundary condition for the fermions is given by the following 
 expression\footnote{We found it convenient to introduce the parameter $\eta_1$
 which takes on the values $\pm 1$ and correspond to the
 choice of spin structure.} \cite{Lindstrom:2002jb}
\beq
 \psi_-^\mu = \eta_1 R^\mu_{\,\,\nu} (X) \psi^\nu_+
\eeq{fermbc} 
 In terms of the fermionic component (\ref{compgrcur}) of the N=1 affine currents, equation 
  (\ref{fermbc})  can be rewritten  as
\beq
 j_-^A = \eta_1 R^A_{\,\,B} (X) j_+^B
\eeq{fermbccurgr}
 where $R^\mu_{\,\,\nu} =l^\mu_A R^A_{\,\,B}r^B_\nu$. In what follows we focus
  on the case when $R^A_{\,\,B}$ is independent of $X$. This is the case usually 
 considered in the study of boundary CFT. 

 We want to understand what kind of restrictions should be imposed on 
 $R^A_{\,\,B}$ for  N=2 supersymmetry to be preserved. 
 In components the manifest on-shell supersymmetry transformations are 
\beq
\left \{ \begin{array}{l}
 \delta_1 X^\mu = - (\epsilon^+_1 \psi_+^\mu + \epsilon_1^- \psi_-^\mu) \\
 \delta_1 \psi_+^\mu = -i \epsilon_1^+ \d_\+ X^\mu + \epsilon_1^- 
 \Gamma^{-\mu}_{\,\,\nu\rho} \psi_-^\rho \psi_+^\nu \\
 \delta_1 \psi_-^\mu = -i\epsilon_1^- \d_= X^\mu - \epsilon_1^+ \Gamma^{-\mu}_{\,\,\nu\rho}
 \psi_-^\rho \psi_+^\nu
\end{array}
\right .
\eeq{n11aa}
 and the nonmanifest supersymmetry transformations (\ref{secsupfl}) are 
\beq
\left \{\begin{array}{l}
 \delta_2 X^\mu =   \epsilon^+_2 \psi_+^\nu J^\mu_{+\nu} 
 + \epsilon_2^- \psi_-^\nu J^\mu_{-\nu} \\
 \delta_2 \psi_+^\mu = -i \epsilon_2^+ \d_\+ X^\nu J^\mu_{+\nu} - \epsilon_2^- 
 J^\mu_{-\sigma}\Gamma^{-\sigma}_{\,\,\nu\rho} \psi_-^\rho \psi_+^\nu +
\epsilon_2^+ J^\mu_{+\nu,\rho} \psi_+^\nu \psi_+^\rho + \epsilon_2^- J^\mu_{-\nu,\rho}
 \psi_-^\nu \psi_+^\rho \\
 \delta_2 \psi_-^\mu = -i\epsilon_2^- \d_= X^\nu J^\mu_{-\nu} 
 + \epsilon_2^+ J^\mu_{+\sigma}\Gamma^{-\sigma}_{\,\,\nu\rho}
 \psi_-^\rho \psi_+^\nu +\epsilon_2^+ J^\mu_{+\nu,\rho} \psi_+^\nu \psi_-^\rho
 + \epsilon_2^- J^\mu_{-\nu,\rho}\psi_-^\nu \psi_-^\rho .
\end{array}
\right .
\eeq{n11bb}
 Starting from the fermionic ansatz (\ref{fermbc}) and applying both supersymmetry 
 transformations, (\ref{n11aa}) and (\ref{n11bb}) we should get the bosonic boundary
  conditions.
 The result of the first transformation is
\beq
 \d_= X^\mu - R^\mu_{\,\,\nu}\d_\+ X^\nu + 2i (P^\sigma_{\,\,\gamma} \nabla_\sigma
 R^\mu_{\,\,\nu} + P^\mu_{\,\,\rho} g^{\rho\delta} H_{\delta\sigma\gamma}
 R^\sigma_{\,\,\nu})\psi_+^\gamma \psi_+^\nu =0
\eeq{firstsuH}
 where $\epsilon_1^+=\eta_1\epsilon_1^-$.  
 The second supersymmetry gives
\ber
\nonumber
&& \d_= X^\mu + (\eta_1\eta_2) J^\mu_{-\lambda} R^\lambda_{\,\,\sigma} J^\sigma_{+\nu}
\d_\+ X^\nu + i \left [ (\eta_1\eta_2) J^\mu_{-\lambda}\nabla^{(-)}_\rho R^\mu_{\,\,\nu} 
 J^\rho_{+\gamma}+\right . \\
&&\left . + (\eta_1\eta_2) J^\mu_{-\lambda} R^\lambda_{\,\,\sigma} J^\sigma_{+\rho} H^\rho_{\,\,\nu\gamma}
 + J^\mu_{-\lambda} \nabla^{(+)}_\rho R^\lambda_{\,\,\nu} J^\rho_{-\sigma} R^\sigma_{\,\,\gamma}
 - H^\mu_{\,\,\rho\sigma} R^\sigma_{\,\,\gamma} R^\rho_{\,\,\nu}\right ]\psi_+^\gamma \psi_+^\nu = 0 
\eer{seconsusyH}
 where $\epsilon_2^+=\eta_2\epsilon_2^-$ and we have used the property (\ref{nablJH}). 
 The boundary conditions (\ref{firstsuH}) and (\ref{seconsusyH}) should be equivalent. 
 Starting from the X-part we get the condition
\beq
 J^\mu_{-\nu} R^\nu_{\,\,\lambda} =  (\eta_1\eta_2) 
 R^\mu_{\,\,\nu} J_{+\lambda}^\nu .
\eeq{biherjRjR}
 In analogy with the K\"ahler case we use the notation ``A-type'' condition when $\eta_1\eta_2=-1$
 and ``B-type'' when 
 $\eta_1\eta_2=1$.

 Using the (\ref{biherjRjR}) the equation (\ref{seconsusyH}) is rewritten as
\ber
\nonumber
&& 
\d_= X^\mu - R^\mu_{\,\,\nu}
\d_\+ X^\nu + i \left [ (\eta_1\eta_2) J^\mu_{-\lambda}\nabla^{(-)}_\rho R^\mu_{\,\,\nu} 
 J^\rho_{+\gamma}+\right . \\
&&\left . +
 (\eta_1\eta_2) J^\mu_{-\lambda} \nabla^{(+)}_\rho R^\lambda_{\,\,\nu} R^\rho_{\,\,\sigma} J^\sigma_{+\gamma}
 - R^\mu_{\,\,\lambda} H^\lambda_{\,\,\nu\gamma} 
   - H^\mu_{\,\,\rho\sigma} R^\sigma_{\,\,\gamma} R^\rho_{\,\,\nu}\right ]\psi_+^\gamma \psi_+^\nu = 0 
\eer{simpBhbbc}
 Using (\ref{inegrabtroB}) we further rewrite (\ref{simpBhbbc}) as 
\beq
 \d_= X^\mu - R^\mu_{\,\,\nu}
\d_\+ X^\nu + 2i J_{+\gamma}^\sigma J_{+\nu}^\lambda \left ( P^\rho_{\,\,\sigma}
 \nabla_\rho R^\mu_{\,\,\lambda} + P^\mu_{\,\,\phi} H^\phi_{\,\,\rho\sigma} R^\rho_{\,\,\lambda}
 \right ) \psi_+^\gamma \psi_+^\nu =0
\eeq{simptorHHH}
Comparing the two-fermion terms of (\ref{firstsuH}) and (\ref{simptorHHH}) 
 we get
\beq
 P^\sigma_{\,\,[\gamma|} \nabla_\sigma
 R^\mu_{\,\,|\nu]} + P^\mu_{\,\,\rho}  H_{\rho\sigma[\gamma}
 R^\sigma_{\,\,\nu]}
-J_{+[\gamma}^\sigma J_{+\nu]}^\lambda \left ( P^\rho_{\,\,\sigma}
 \nabla_\rho R^\mu_{\,\,\lambda} + P^\mu_{\,\,\phi} H^\phi_{\,\,\rho\sigma} R^\rho_{\,\,\lambda}
 \right ) = 0 .
\eeq{twopferH}
 Using the projectors $\Omega^+_{\pm}=1/2(I\pm iJ_+)$ we rewrite the above 
 condition as
\beq
\Omega_{\pm[\gamma}^{+\sigma} \Omega_{\pm\nu]}^{+\lambda} \left ( P^\rho_{\,\,\sigma}
 \nabla_\rho R^\mu_{\,\,\lambda} + P^\mu_{\,\,\phi} H^\phi_{\,\,\rho\sigma} R^\rho_{\,\,\lambda}
 \right ) = 0
\eeq{condomHH1}
 Thus, using only the supersymmetry transformations we have obtained two conditions  on 
 $R^\mu_{\,\,\nu}$, (\ref{biherjRjR}) and (\ref{condomHH1}). In turn the condition 
 (\ref{biherjRjR}) implies
\beq
 J^A_{\,\,C} R^C_{\,\,B} = (\eta_1\eta_2) R^A_{\,\,C} J^C_{\,\,B} 
\eeq{condRJRgrp}  
 Assuming that $R^A_{\,\,B}$ is constant (\ref{condomHH1}) implies
\beq
 R^A_{\,\,M} f_{BN}^{\,\,\,\,\,\,\,\,M} +  R^C_{\,\,B} R^M_{\,\,N} f_{CM}^{\,\,\,\,\,\,\,\,A}
 = J^S_{\,\,B} J^L_{\,\,N} \left ( R^A_{\,\,M}f_{SL}^{\,\,\,\,\,\,\,\,M}  
 + R^C_{\,\,S} R^M_{\,\,L}f_{CM}^{\,\,\,\,\,\,\,\,A}\right )
\eeq{necondgrpJJ}
 To  better understand the meaning of the conditions (\ref{condRJRgrp}) and (\ref{necondgrpJJ})
 we choose the basis adapted to the Manin triple $({\bf g}, {\bf g_-}, {\bf g}_+)$
  discussed in the previous
 section. For B-type conditions, (\ref{condRJRgrp}) implies
 that $R^a_{\,\,\bar{b}}=R^{\bar{a}}_{\,\,b}=0$. Taking this into account 
 we rewrite the condition (\ref{necondgrpJJ}) as follows
\beq
  R^c_{\,\,b} R^m_{\,\,n} f_{cm}^{\,\,\,\,\,\,a} + R^a_{\,\,m} f_{bn}^{\,\,\,\,\,\,m} = 0
\eeq{condholBt1}
\beq
  R^{\bar{c}}_{\,\,\bar{b}} R^{\bar{m}}_{\,\,\bar{n}} 
 f_{\bar{c}\bar{m}}^{\,\,\,\,\,\,\bar{a}} + R^{\bar{a}}_{\,\,\bar{m}} 
 f_{\bar{b}\bar{n}}^{\,\,\,\,\,\,\bar{m}} = 0 .
\eeq{condholBt2}
 We conclude that $R^a_{\,\,b}$ is a Lie algebra automorphism for ${\bf g_+}$
 (more presicely, $[T_a, T_b] = - f_{ab}^{\,\,\,\,\,\,c} T_c$) and that $R^{\bar{a}}_{\,\,\bar{b}}$
  is a Lie algebra automorphism for ${\bf g_-}$. 

 For the A-type boundary conditions,  (\ref{condRJRgrp}) yields 
 that $R^a_{\,\,b}=R^{\bar{a}}_{\,\,\bar{b}}=0$.  Taking this  into account 
 we rewrite the condition (\ref{necondgrpJJ}) as
\beq
    R^{\bar{c}}_{\,\,b} R^{\bar{m}}_{\,\,n} 
 f_{\bar{c}\bar{m}}^{\,\,\,\,\,\,\bar{a}} + R^{\bar{a}}_{\,\,m} 
 f_{bn}^{\,\,\,\,\,\,m} = 0 ,
\eeq{condholAt1}
\beq
  R^{c}_{\,\,\bar{b}} R^{m}_{\,\,\bar{n}} 
 f_{cm}^{\,\,\,\,\,\,a} + R^{a}_{\,\,\bar{m}} 
 f_{\bar{b}\bar{n}}^{\,\,\,\,\,\,\bar{m}} = 0 .
\eeq{condholAt2}
 We conclude that $R^{\bar{a}}_{\,\,b}$ is a Lie algebra homomorphism from ${\bf g_+}$ 
 to ${\bf g_-}$ and  $R^{a}_{\,\,\bar{b}}$ is a Lie algebra homomorphism from ${\bf g_-}$ 
 to ${\bf g_+}$. 

 We summarize the results. For the B-type supersymmetry we have the following boundary 
 conditions
\beq
 j_-^a = \eta_1 R^a_{\,\,b} j_+^b,\,\,\,\,\,\,\,\,\,\,\,
k_=^a =R^a_{\,\,b} k_\+^b,\,\,\,\,\,\,\,\,\,\,\,
j_-^{\bar{a}} = \eta_1 R^{\bar{a}}_{\,\,\bar{b}} j_+^{\bar{b}},\,\,\,\,\,\,\,\,\,\,\,
k_=^{\bar{a}} =R^{\bar{a}}_{\,\,\bar{b}} k_\+^{\bar{b}},
\eeq{bcBsusyres}  
 where $R^a_{\,\,b}$ is a Lie algebra automorphism for ${\bf g_-}$ and  $R^{\bar{a}}_{\,\,\bar{b}}$ 
 is a Lie algebra automorphism for ${\bf g_+}$. For the A-type supersymmetry 
  the boundary  conditions are
\beq
 j_-^a = \eta_1 R^a_{\,\,\bar{b}} j_+^{\bar{b}},\,\,\,\,\,\,\,\,\,\,\,
k_=^a =R^a_{\,\,\bar{b}} k_\+^{\bar{b}},\,\,\,\,\,\,\,\,\,\,\,
j_-^{\bar{a}} = \eta_1 R^{\bar{a}}_{\,\,b} j_+^{b},\,\,\,\,\,\,\,\,\,\,\,
k_=^{\bar{a}} =R^{\bar{a}}_{\,\,b} k_\+^{b},
\eeq{AcBsusyres}  
 where $R^a_{\,\,\bar{b}}$ is a Lie algebra homomorphism from ${\bf g_-}$ to ${\bf g_+}$ 
 and  $R^{\bar{a}}_{\,\,b}$  is a Lie algebra homomorphism from ${\bf g_+}$ to ${\bf g_-}$. 
 It is important to stress that the requirement of conformal invariance does
 not enter here.  In our derivation we used only the supersymmetry 
 transformations and the assumption that $R^A_{\,\,B}$ is a constant. 

 In the above derivation we analysed the problem  first in terms of $X^\mu$ and $\psi^\mu_\pm$
 and then expressed the results in terms of the affine currents. 
 Of course we could first rewrite the supersymmetry transformations in terms of the affine currents
 and then the analyse the boundary condition for the affine currents. The result would not change.
 For the sake of completeness we also record the supersymmetry transformations
 in terms of affine currents. 

The manifest supersymmetry
\beq
 \delta_1 {\cal J}^A_\pm = i \epsilon_1^+ Q_+ {\cal J}_{\pm}^A +i \epsilon_1^- Q_- {\cal J}_{\pm}^A  
\eeq{manforcuR} 
 with $Q_{\pm}$ defined in (\ref{alg}) can be written in the components as follows
\beq
\left \{ \begin{array}{l}
 \delta_1 j_\pm^A = - i\epsilon_1^\pm k^A_{\pp}   \\
 \delta_1 k_{\pp}^A = - \epsilon^\pm_1 \d_{\pp} j^A_\pm
\end{array}
\right .
\eeq{currmansusy}
 where we have used the equations of motion. Using the definition (\ref{fefgroupcur})
 of N=1 affine currents we can write the nonmanifest supersymmetry (\ref{secsupfl}) (on-shell) as 
\beq
 \delta_2 {\cal J}_{\pm}^A = -\epsilon^\pm_2 J^A_{\,\,B} D_{\pm} {\cal J}^B_{\pm}
 \mp \epsilon^\pm_2 f_{MK}^{\,\,\,\,\,\,\,\,A} J^K_{\,\,B} {\cal J}^B_\pm {\cal J}^M_\pm .
\eeq{nonmancursusy}
In components the transformation (\ref{nonmancursusy}) become
\beq
\left \{ \begin{array}{l}
 \delta_2 j_\pm^A = - i\epsilon_2^\pm  J^A_{\,\,B} k^B_{\pp} \mp \epsilon^\pm_2 f_{MK}^{\,\,\,\,\,\,\,\,A} 
 J^K_{\,\,B} j^B_\pm j^M_\pm   \\
 \delta_2 k_{\pp}^A =  \epsilon^\pm_2 J^A_{\,\,B} \d_{\pp} j^B_\pm \mp \epsilon_2^\pm
  f_{K[B}^{\,\,\,\,\,\,\,\,\,\,\,\,\,A} J^K_{\,\,M]} k_{\pp}^M j_{\pp}^B 
\end{array}
\right .
\eeq{currnonmansusy}
 Now starting from $j_-^A = \eta_1 R^A_{\,\,B} j_+^B$ and using the transformations (\ref{currmansusy})
 and (\ref{currnonmansusy}) we easily rederive the previous results (\ref{condRJRgrp}) 
 and (\ref{necondgrpJJ}). 

In fact the transformations (\ref{currmansusy}) and (\ref{currnonmansusy}) can
  be ``complexified'' in the Manin basis. Introducing $\delta=\delta_1 + \delta_2$ 
 and $\epsilon^\alpha = \epsilon_1^\alpha + i \epsilon_2^\alpha$ we can write the transformation 
 as follows
\beq
\left \{ \begin{array}{l}
 \delta j_\pm^a = - i\epsilon^\pm  k^a_{\pp} \pm \frac{1}{2}(\epsilon^\pm -\bar{\epsilon}^\pm)
   f_{bc}^{\,\,\,\,\,\,a} j_{\pm}^b j_\pm^c \\ 
 \delta k_{\pp}^a =  -\bar{\epsilon}^\pm  \d_{\pp} j^a_\pm \pm (\epsilon^\pm - \bar{\epsilon}^\pm)
  f_{bc}^{\,\,\,\,\,\,a} k_{\pp}^c j_\pm^b
\end{array}
\right .
\eeq{cursusycompl}
 (with a similar expression for the $\bar{a}$-part).

\Section{N=2 supreconformal boundary conditions}
\label{s:conformal}

In this section we incorporate the requirement of 
 conformal invariance into the boundary  conditions. 
  We derive the N=2 superconformal boundary conditions by imposing appropriate
 boundary  conditions on the conserved (2,2) currents ($T_{\pm\pm}$, $G_{\pm}^1$, $G_{\pm}^2$, $J_{\pm}$).
 The N=1 superfield and component forms of these currents can be found in 
 \cite{Lindstrom:2002jb}. Here we  present them in terms
 of the  fermionic and bosonic affine  currents, $j^A_{\pm}$ and $k^A_{\pp}$. 
 Using results from \cite{Lindstrom:2002jb}
 and the definitions (\ref{compgrcur}) it is a straightforward exercise to write the N=2
  currents as
\beq 
T_{++} = k_{\+}^A \eta_{AB} k_{\+}^B + i j_+^A \eta_{AB} \d_\+
 j^B_+ + i k_{\+}^A j^B_+ j^C_+ f_{ABC} , 
\eeq{comp3} 
\beq 
T_{--} =  k_{=}^A \eta_{AB} k_{=}^B + i j_-^A \eta_{AB} \d_=
 j^B_- - i k_{=}^A j^B_- j^C_- f_{ABC} , 
\eeq{comp4N2H}
 \beq 
G^1_{+} =  j_+^A \eta_{AB} k_{\+}^B + \frac{i}{3} j_+^A j_+^B j_+^C f_{ABC} , 
\eeq{comp1N2H} 
\beq 
G^1_{-} = j_-^A \eta_{AB} k_{=}^B - \frac{i}{3} j_-^A j_-^B j_-^C f_{ABC} , 
\eeq{comp2N2H} 
\beq
 G^2_{+} =  j_+^A J_{AB} k_{\+}^B  ,
\eeq{compN2Ha}
\beq
 G^2_{-} = j_-^A J_{AB} k_{=}^B ,
\eeq{compN2Hb}
\beq
 J_+ = j_+^A j_+^B J_{AB},\,\,\,\,\,\,\,\,\,\,\,\,\,\,\,\,
 J_{-} = j_-^A j_-^B J_{AB} .
\eeq{compN2Hcrsym}
 We define the following linear combinations of $G^i_{\pm}$
\beq
 {\cal G}_\pm = \frac{1}{2} (G^1_\pm + i G^2_\pm),\,\,\,\,\,\,\,\,\,\,\,\,\,\,\,\,\,
 \bar{\cal G}_\pm = \frac{1}{2} (G^1_\pm - i G^2_\pm)
\eeq{defcrulyG}
 Using the properties of Manin triple $({\bf g}, {\bf g_-}, {\bf g_+})$ we write 
 ${\cal G}_\pm$ and $\bar{\cal G}_\pm$ as follows 
\beq
  {\cal G}_{\pm} = j_\pm^a \eta_{a\bar{b}} k_{\pp}^{\bar{b}} \pm \frac{i}{2}
  j_\pm^a j_\pm^{\bar{b}} j_\pm^{\bar{c}}f_{a \bar{b}\bar{c}}  \pm \frac{i}{2}
  j_\pm^a j_\pm^{b} j_\pm^{\bar{c}}
 f_{a b\bar{c}} ,
\eeq{newG}
\beq
  \bar{\cal G}_{\pm} = j_\pm^{\bar{a}} \eta_{\bar{a}b} k_{\pp}^{b} \mp \frac{i}{2} j_\pm^a 
 j_\pm^{b} j_\pm^{\bar{c}} f_{ab\bar{c}}  \mp \frac{i}{2}
  j_\pm^a j_\pm^{\bar{b}} j_\pm^{\bar{c}}f_{a \bar{b}\bar{c}} .
\eeq{newGbar}
 We see that once we have a Lie algebra ${\bf g}$ with invariant inner product $\eta$ (given by 
  $\eta_{AB}$) and a Manin triple $({\bf g}, {\bf g_-}, {\bf g_+})$ defined with respect to $\eta$ we
 may define the N=2 currents (\ref{comp3}), (\ref{comp4N2H}), (\ref{compN2Hcrsym}), 
 (\ref{newG}) and (\ref{newGbar}).    
  In fact they will obey the correct N=2 alegbra\footnote{One should keep 
 in mind that there are two sets of bosonic affine currents which differ by the two-fermion
 term.}, \cite{Getzler:py} 
 and \cite{Getzler:fs}. 

To ensure N=2 superconformal symmetry on the boundary 
we have to impose the following conditions\footnote{Classically these conditions make 
 sense only on-shell
 since the currents are defined modulo the equations
 of motion.} on the 
currents (\ref{comp3})--(\ref{compN2Hcrsym}), 
\beq
 T_{++}-T_{--}=0,\,\,\,\,\,\,\,\,\,\,\,\,\,\,\,\,\,\,
 G^1_+-\eta_1 G^1_- =0,
\eeq{curbounN1}
\beq
 G_+^2 - \eta_2 G^2_- =0, \,\,\,\,\,\,\,\,\,\,\,\,\,\,\,\,\,\,
 J_+ - (\eta_1 \eta_2) J_- =0. 
\eeq{curbounN2H}
 The conditions (\ref{curbounN1}) ensure N=1 superconformal invariance. 
 Starting from the ansatz $j_-^A=\eta_1 R^A_{\,\,B} j_+^B$ with constant $R^A_{\,\,B}$
 and solving the conditions (\ref{curbounN1}) we derive the bosonic counterpart
\beq
 k_=^A = R^A_{\,\,B} k_\+^B,
\eeq{bosnbcN1}
 together with the additional properties
\beq
 R^C_{\,\,A} \eta_{CD} R^D_{\,\,B} = \eta_{AB},\,\,\,\,\,\,\,\,\,\,\,\,\,\,\,
 f_{ABC} + R^L_{\,\,A} R^M_{\,\,B} R^N_{\,\,C} f_{LMN} = 0
\eeq{resultN1}
 Thus N=1 superconformal invarince implies that $R^A_{\,\,B}$ should be 
 a Lie algebra automorphism. Solving the conditions (\ref{curbounN2H}) we arrive 
 at the condition
\beq
 J^A_{\,\,C} R^C_{\,\,B} = (\eta_1\eta_2) R^A_{\,\,C} J^C_{\,\,B}.
\eeq{N2invdh}
 As one would expect the conditions (\ref{resultN1}) and (\ref{N2invdh}) are stronger
 than the conditions (\ref{condRJRgrp}) and 
 (\ref{necondgrpJJ}) which come  from the N=2 supersymmetry alone. The difference 
 is the condition $R^C_{\,\,A} \eta_{CD} R^D_{\,\,B} = \eta_{AB}$. Thus adding
  this condition to  (\ref{condRJRgrp}) and 
 (\ref{necondgrpJJ}) we recover (\ref{resultN1}) 
 and (\ref{N2invdh}).

The conserved currents $J_{\pm}$ generate two R-rotations
 which act trivially on the bosonic fields but non-trivially on the fermions.
 Because of the boundary condition $J_{+}-(\eta_1 \eta_2) J_{-}=0$ only one 
 combination of these R-rotations survives as a symmetry in the presence of
 a boundary. Thus for the B-type 
 we have the following R-symmetry
\beq
\left \{ \begin{array}{l}
 j_+^A \rightarrow \cos \alpha \,\,j_+^A + \sin \alpha \,\, J^A_{\,\,B} j_+^B \\
 j_-^A \rightarrow \cos \alpha \,\,j_-^A + \sin \alpha \,\, J^A_{\,\,B} j_-^B 
\end{array} \right .
 \eeq{RsymBtyK} 
 and for the A-type
\beq
\left \{ \begin{array}{l}
 j_+^A \rightarrow \cos \alpha \,\,j_+^A + \sin \alpha \,\, J^A_{\,\,B} j_+^B \\
 j_-^A \rightarrow \cos \alpha \,\,j_-^A - \sin \alpha \,\, J^A_{\,\,B} j_-^B 
\end{array} \right .
 \eeq{RsymAtyK} 
In the Manin basis these rotations are  $(j_\pm^a \rightarrow e^{i\alpha}
 j_\pm^a,\,\,\,\,j_\pm^{\bar{a}} \rightarrow e^{-i\alpha} j_\pm^{\bar{a}})$ and
  $(j_\pm^a \rightarrow e^{\pm i\alpha}
 j_\pm^a,\,\,\,\,j_\pm^{\bar{a}} \rightarrow e^{\mp i\alpha} j_\pm^{\bar{a}})$
 respectively.

\section{Summary and discussion}
\label{s:end}

 We consider a WZW model defined over a connected Lie group ${\cal G}$, such 
  such that its Lie algebra ${\bf g}$ comes equipped with 
 a symmetric ad-invariant nondegenerate bilinear form $(T_A,T_B)=\eta_{AB}$ and can
 be decomposed into a pair of maximally isotropic subalgebras ${\bf g_-}$, ${\bf g_+}$
 with respect to $\eta$ and ${\bf g}$ as a vector space is the direct sum of ${\bf g_-}$
 and ${\bf g_+}$. This ordered triple of algebras $({\bf g}, {\bf g_-}, {\bf g_+})$
 is called a Manin triple. It is easy to  see that  the dimensions of sublagebras
 ${\bf g_-}$, ${\bf g_+}$ are equal and that the bases $\{T_a \}$, $\{ T^a \}$
 may be choosen such that 
\beq
 (T_a, T_b)=0,\,\,\,\,\,\,\,\,\,\,\,\,
 (T_a, T^b)=\delta_{\,\,a}^{b},\,\,\,\,\,\,\,\,\,\,\,\,
 (T^a, T^b)=0 .
\eeq{basisMt}
 The algebraic structure of ${\bf g}$ is  completely determined by the structures
 of the maximal isotropic subalgebras in the basis (\ref{basisMt}) 
\beq
 [T_a, T_b ]= f_{ab}^{\,\,\,\,\,\,c}T_c,\,\,\,\,\,\,\,\,\,\,\,\,
 [T^a, T^b] = f^{ab}_{\,\,\,\,\,\,c} T^c,\,\,\,\,\,\,\,\,\,\,\,\,
 [T_a, T^b] = f_{ca}^{\,\,\,\,\,\,b} T^c + f^{bc}_{\,\,\,\,\,\,a} T_c .
\eeq{fulMMA}
 Introducing the corresponding affine fermionic and bosonic currents $(j_{\pm a}, k_{\pp a})$
 and $(j^a_\pm, k^a_{\pp} )$ we construct the N=(2,2) currents ($T_{\pm\pm}$, $G_{\pm}^1$,
  $G_{\pm}^2$, $J_{\pm}$) as defined in (\ref{comp3}), (\ref{comp4N2H}), (\ref{compN2Hcrsym}), 
 (\ref{newG}) and (\ref{newGbar}) correspondently. In the presence of 
 a boundary, appropriate gluing conditions should be imposed on the affine 
 currents. There are two ways of gluing the fermionic currents: the first is
\beq
 j^a_- = \eta_1 R^a_{\,\,b} j_+^b ,
\eeq{Btypecon}
 which we call the B-type and the second is
\beq
 j^a_- = \eta_1 R^{ab} j_{+b} ,
\eeq{Atypecon}
 which we call the A-type. To preserve half of the bulk supersymmetry
 the gluing matrix should satisfy the following: $R^a_{\,\,b} \in Aut({\bf g_-})$
 for the B-type and $R^{ab} \in Hom({\bf g_-}, {\bf g_+})$ for the A-type. 
 Gluing conditions must also be imposed on the bosonic affine currents.
 In addition conformal invariance requires $R$ to preserve the form $\eta$.
 Therefore,  for both A- and B-types, superconformal boundary conditions require $R \in Aut({\bf g})$. 

 It is easy to give the example of the B-type brane with $R=I$ which 
 would correspond the branes localized along conjugacy classes \cite{Alekseev:1998mc}. 
  However it is a bit problematic to give a simple example of the A-type brane
 since it would depend on the definition of the form $\eta$. We hope to come 
 back to these examples elsewhere.

\bigskip 
 
\bigskip

{\bf Acknowledgements}: 
 UL acknowledges support in part by EU contract  
 HPNR-CT-2000-0122 and by VR grant 650-1998368. 
 MZ acknowledges support in part by EU contract
  HPRN-CT-2002-00325.

\appendix 
 
\section{(1,1) supersymmetry} 
\label{a:11susy} 

 In this we collect our conventions on the N=1 supersymmetry which we use through the text.
 
We deal with real (Majorana) two-component spinors $\psi^\alpha= 
 (\psi^+, \psi^-)$. Spinor indices are raised and lowered by the second-rank antisymmetric 
 symbol $C_{\alpha\beta}$, which defines the spinor inner product: 
\beq 
 C_{\alpha\beta}=-C_{\beta\alpha}=-C^{\alpha\beta},\,\,\,\,\, C_{+-}=i,\,\,\,\,\,
\psi_\alpha =\psi^\beta C_{\beta\alpha},\,\,\,\,\, \psi^\alpha= C^{\alpha\beta} \psi_\beta.
\eeq{Cdef} 
Throughout the paper we use  $(\+,=)$ as worldsheet indices, and $(+,-)$ as two-dimensional spinor 
indices.  We also use superspace conventions where the pair of spinor 
coordinates of the two-dimensional superspace are labelled $\th^{\pm}$, 
and the covariant derivatives $D_\pm$ and supersymmetry generators 
$Q_\pm$ satisfy 
\ber 
D^2_+ &=&i\d_\+, \quad 
D^2_- =i\d_= \quad \{D_+,D_-\}=0\cr 
Q_\pm &=& iD_\pm+ 2\th^{\pm}\d_{\pp} 
\eer{alg} 
where $\d_{\pp}=\partial_0\pm\partial_1$.  In terms of the covariant 
derivatives, a supersymmetry transformation of a superfield $\P$ is 
then given by 
\ber 
\delta \P &\equiv & i(\e^+Q_++\e^-Q_-)\P \cr 
&=& -(\e^+D_++\e^-D_-)\P 
+2i(\e^+\th^+\d_\++\e^-\th^-\d_=)\P . 
\eer{tfs} 
The components of a superfield $\P$ are defined via projections as 
follows, 
\ber 
\P|\equiv X, \quad D_\pm\P| \equiv \p_\pm, \quad D_+D_-\P|\equiv F_{+-} 
, 
\eer{comp} 
where a vertical bar denotes ``the $\th =0$ part of ''. 
Thus, in components, the $(1,1)$ supersymmetry transformations are given by 
\beq 
\left \{ \begin{array}{l} 
    \delta X^\mu = - \epsilon^{+} \psi_+^\mu - \epsilon^- \psi_-^\mu \\ 
    \delta \psi_+^\mu =  -i\epsilon^+ \d_{\+}X^\mu + \epsilon^- F^\mu_{+-}\\ 
    \delta \psi_-^\mu  = -i \epsilon^- \d_{=} X^\mu - \epsilon^+ F_{+-}^\mu \\ 
    \delta F^\mu_{+-} = - i \epsilon^+ \d_{\+} \psi_-^\mu + i 
    \epsilon^- \d_- \psi_+^\mu 
\end{array} \right . 
\eeq{compsusytrN1}

\end{document}